\documentclass[%
 reprint,
 amsmath,amssymb,
 aps,prl
]{revtex4-2}

\usepackage[utf8]{inputenc}
\usepackage{amsmath}
\usepackage{bm}
\usepackage{graphicx}
\usepackage{mathrsfs}

\usepackage{hyperref}
\usepackage{color}

\begin{document}

\title{Chemistry of a light impurity in a Bose-Einstein condensate}

\author{Arthur Christianen}
\email{arthur.christianen@mpq.mpg.de}

\author{J. Ignacio Cirac}%

\author{Richard Schmidt}%
\email{richard.schmidt@mpq.mpg.de}
\affiliation{%
Max-Planck-Institut für Quantenoptik, Hans-Kopfermann-Str. 1, D-85748 Garching, Germany
}%
\affiliation{%
Munich Center for Quantum Science and Technology (MCQST), Schellingstraße 4, D-80799 Munich, Germany
}%

\date{\today}
\begin{abstract}
    In ultracold atomic gases, a unique interplay arises between phenomena known from condensed matter physics, few-body physics and chemistry. Similar to an electron in a solid, an impurity in an ultracold gas can get dressed by excitations from the medium, forming a quasiparticle called the polaron. We study how dressing of an impurity leads to a modification of its chemical reactivity. Using a Gaussian state variational method in the frame of the impurity, we demonstrate that three-body correlations lead to an instability of the polaron. This instability is connected to an Efimov resonance, but shifted to smaller interactions by many-body effects, showing that polaron formation stimulates Efimov physics and the associated chemistry.
\end{abstract}

\maketitle

The concept of the polaron, an impurity dressed by excitations from a medium, was introduced by Landau \cite{landau:1933} and has developed into an iconic paradigm in condensed matter physics. Polaron physics is important to describe electrons interacting with phonons in solids \cite{alexandrov:2010} and is therefore crucial to understand the properties of systems such as organic semiconductors \cite{hulea:2006,watanabe:2014}, lead halide perovskites \cite{zhu:2016,miyata:2017},  transition metal oxides \cite{katz:2012,verdi:2017} and high-Tc superconductors \cite{salje:2005,zhang:2019}. Importantly, studying polaron problems can give insight into fundamental theoretical questions relating to the general setting of small subsystems interacting with a larger environment. 

A unique window into polaron physics is provided by experiments on ultracold atoms. Here radiofrequency spectroscopy can be used to directly probe both Fermi polarons \cite{schirotzek:2009}, where the medium is a Fermi sea, and Bose polarons, where the medium is a Bose-Einstein condensate (BEC)  \cite{jorgensen:2016,hu:2016,yan:2019}. However, in BECs also chemical reactions involving the impurity can occur, in particular three-body recombination. From the perspective of chemistry, polaron formation is similar to the process of solvation, which can drastically change the chemical properties of the solute. Therefore the question naturally arises how polaron formation and chemical reactions can affect each other.

Three-body recombination due to decay into deeply bound states is an important subject in cold atomic systems, since it is a major factor limiting their lifetime. Moreover, recombination is at the core of research in few-body physics, since it provides a direct tool to study universal bound state physics, in particular the Efimov effect \cite{kraemer:2006,pires:2014,tung:2014}. Efimov showed that there exists an infinite series of universal three-body bound states close to a Feshbach resonance \cite{efimov:1970,naidon:2017}. Also larger bound states of Efimov character (Efimov clusters) have been predicted to exist \cite{vonstecher:2011}, and have been observed experimentally through four- and five-body recombination \cite{ferlaino:2009,zenesini:2013}. 

In this work we theoretically address the question how polaron formation affects resonant three-body recombination due to the Efimov effect. We show that also the reverse effect is important, since the three-body correlations yielding Efimov physics strongly affect the polaron. So far, the simultaneous description of both the many-body physics of polaron cloud formation, which can lead to a large number of excitations surrounding the impurity \cite{shchadilova:2016,guenther:2021}, and the necessary correlations to describe impurity-mediated interactions leading to the Efimov effect \cite{levinsen:2015,yoshida:2018} remains an open challenge. We tackle this problem by using a variational method with a Gaussian State Ansatz \cite{shi:2017} in the reference frame of the impurity. We focus on the challenging case of light impurities, where the Efimov effect is most prominent\cite{pires:2014,tung:2014,naidon:2017,sun:2017} due to a large impurity kinetic energy. 

We find that beyond a critical scattering length $a^{\ast}$, the polaron is unstable towards Efimov cluster formation through two- and three-body scattering processes. The scattering length $a^{\ast}$ is connected to the Efimov scattering length $a_{-}$ in the low density limit and can be interpreted as a many-body shifted Efimov resonance. As an intriguing example of chemistry in a quantum medium,  $|a^{\ast}|$ becomes smaller as the density of the background BEC increases, due to the polaron cloud participating in the formation of Efimov clusters, leading to polaron-enhanced chemical reactions.

\emph{Model}.- We consider an impurity of mass $M$ immersed in a three dimensional, infinite, weakly interacting BEC at zero temperature. The BEC has density $n_0$ and consists of bosons of mass $m$, with interboson and impurity-boson scattering lengths $a_B$ and $a$, and associated coupling constants $g_B$ and $g$.  We use a single channel Hamiltonian with contact interactions, which is assumed to be valid close to a broad Feshbach resonance:

\begin{multline}
\hat{\mathcal{H}}_0=\int \frac{ d^3k}{(2\pi)^3} \frac{\bm{k}^2}{2m} \hat{a}^\dagger_{\bm{k}} \hat{a}_{\bm{k}}+ \frac{\hat{\bm{P}}^2}{2M}+ g \int d^3r \delta(\bm{r}-\hat{\bm{R}}) \hat{a}^{\dagger}_{\bm{r}} \hat{a}_{\bm{r}} \\ + \frac{g_{B}}{2} \int \int d^3r d^3r' \delta(\bm{r}-\bm{r'}) \hat{a}^{\dagger}_{\bm{r'}} \hat{a}^{\dagger}_{\bm{r}} \hat{a}_{\bm{r'}} \hat{a}_{\bm{r}} .
\end{multline}

 The operators $\hat{a}^{\dagger}_{\bm{k}}$ are bosonic creation operators, and the impurity is described in first quantization with operators $\hat{\bm{P}}$ and $\hat{\bm{R}}$. We now apply the Bogoliubov approximation to describe the weakly repulsive BEC, and introduce the Bogoliubov excitation operators $\hat{b}^{\dagger}_{\bm{k}}$, $\hat{b}_{\bm{k}}$ with dispersion $\omega_k$ in the typical way \cite{shchadilova:2016,ownpaper}. We move to the reference frame of the impurity using the Lee-Low-Pines transformation \cite{lee:1953} and set the total momentum of the system to zero. This gives the extended Fröhlich Hamiltonian introduced in Refs.\ \cite{rath:2013,shchadilova:2016}:

\begin{multline}\label{eq:Hamiltonian}
\hat{\mathcal{H}}= \int_{\bm{k}} \ (\omega_k+\frac{k^2}{2 M }) \hat{b}^\dagger_{\bm{k}} \hat{b}_{\bm{k}} + \ \int_{\bm{k}}  \int_{\bm{k'}} \frac{\bm{k} \cdot \bm{k'} \ \hat{b}^\dagger_{\bm{k}} \hat{b}^\dagger_{\bm{k'}} \hat{b}_{\bm{k}} \hat{b}_{\bm{k'}}}{2M} \\
+g n_0  + g \sqrt{n_0} \int_{\bm{k}} \ W_k [\hat{b}^{\dagger}_{\bm{k}} +\hat{b}_{-\bm{k}}]+  \\
+ g \int_{\bm{k}}\int_{\bm{k'}} \ \big[V^{(1)}_{k,k'} \hat{b}^{\dagger}_{\bm{k}} \hat{b}_{\bm{k'}}+\frac{V^{(2)}_{k,k'}}{2}  (\hat{b}^{\dagger}_{\bm{k}} \hat{b}^{\dagger}_{\bm{k'}}+\hat{b}_{-\bm{k}} \hat{b}_{-\bm{k'}})\big].
\end{multline}
Definitions of $\omega_k$, $W_k$,$V^{(1)}_{k,k'}$,  and $V^{(2)}_{k,k'}$ are given in Refs. \cite{shchadilova:2016,ownpaper}. We use the following notation for the integrals: $\int_{\bm{k}} \equiv \int_0^{\Lambda}\int_0^{\Lambda}\int_0^{\Lambda}\frac{d^3k}{(2\pi)^3}$. Here $\Lambda$ is the UV cutoff that regularizes the impurity-boson interaction and serves as the three-body parameter. It is proportional to the inverse of the Van der Waals-length of the true scattering potential and can be fixed by matching the scattering length of the first Efimov resonance $a_{-}$ to the experimental value for the system of interest.

\begin{figure}[!tp]
    \centering
\includegraphics[]{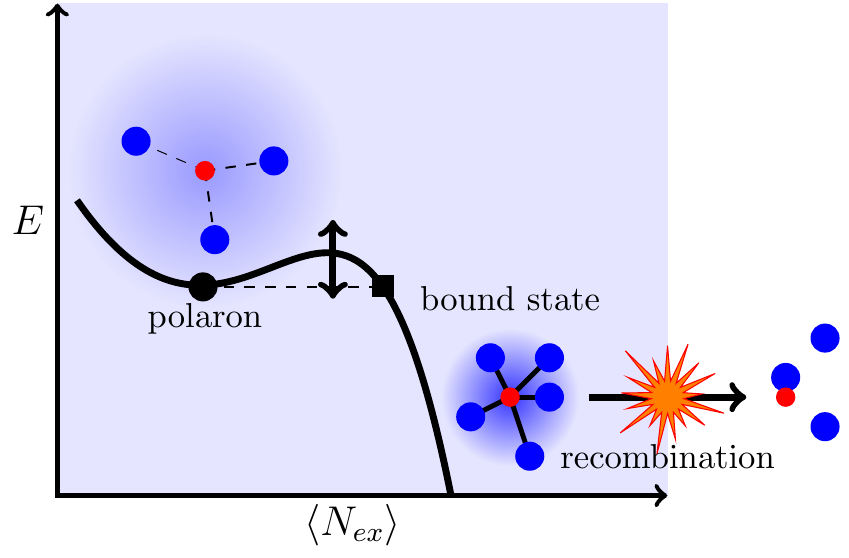}
    \caption{(Color online) Illustration of the energy $E$ of an impurity immersed in a BEC, as a function of the number of excitations $\langle \hat{N}_{ex} \rangle$ surrounding it. There is a local minimum corresponding to the polaron at small numbers of excitations. For large particle numbers bound states can be formed, leading to a much lower energy. The height of the barrier protecting the local minimum is dependent on the density of the BEC and the scattering length. The barrier disappears at the density-dependent critical scattering length $a^{\ast}$. Formation of clusters experimentally leads to recombination. The black circle and square are of relevance for Fig.\ \ref{fig:polaron_partcnumb}.}
    \label{fig:cartoon}
\end{figure}

We describe the bosonic excitations of the BEC by a Gaussian state \cite{shi:2017} variational Ansatz. This allows for an arbitrary number of excitations and pairwise interboson correlations, which by virtue of the Lee-Low-Pines transformation also represent the three-body correlations between two bosons and the impurity. The Gaussian state Ansatz can be written as: 

\begin{equation}
   | \psi \rangle = e^{\bm{\hat{\Psi}}^{\dagger}\Sigma^z \bm{\Phi}} e^{\frac{i}{2} \hat{\bm{\Psi}}^{\dagger} \Xi \hat{\bm{\Psi}}} |BEC \rangle,
\end{equation}
with the Nambu vector $\hat{\Psi}_{\bm{k}}=(\hat{b}_{\bm{k}}^{\dagger}, \hat{b}_{\bm{k}})^T$, the coherent displacement $\bm{\Phi}=\langle \bm{\hat{\Psi}} \rangle$, the correlation matrix $\Xi$, and $\Sigma_{\bm{k,k'}}^z=\delta_{\bm{k,k'}}\sigma^z $, where $\sigma^z$ is the Pauli z-matrix.
Matrix multiplications imply integration over $\bm{k}$.
Both $\bm{\Phi}$ and the elements of $\Xi$ are the variational parameters. We optimize them using imaginary time evolution \cite{shi:2017}.

We consider an exemplary mass ratio $M/m=6/133$ of a $^6$Li impurity in a BEC of $^{133}$Cs atoms.

\emph{Qualitative picture.-} In Fig.\ \ref{fig:cartoon} we schematically plot the energy landscape of the extended Fröhlich Hamiltonian for fixed $n_0$ and fixed negative $a$, as a function of the number of excitations $\langle \hat{N}_{ex} \rangle =\langle \sum_{\bm{k}} \hat{b}^{\dagger}_{\bm{k}} \hat{b}_{\bm{k}} \rangle$ forming the dressing cloud of the impurity. Even in absence of two- and three-body bound states, so for $|a|<|a_{-}|$, we find that the polaron is not the ground state but a local minimum for a large range of interaction strengths. The global minimum of the energy landscape represents a many-body bound state, a large Efimov cluster that forms due to a ``cooperative binding effect" \cite{ownpaper}. At a density-dependent critical scattering length $a^{\ast}$, the polaron breaks down because the barrier protecting the local minimum disappears. This breakdown should be experimentally observable as enhanced three-body recombination, similar to studies of the Efimov effect \cite{tung:2014,pires:2014} in the low density region.

\emph{Formation of Efimov clusters.-} First we consider the right part of Fig.\ \ref{fig:cartoon} and show that the global minimum of our Hamiltonian is indeed a many-body bound state already for small negative scattering lengths. Since the bound states are much more tightly bound than the typical interparticle distance in the BEC, the effect of the medium on the energy of these bound states is relatively small. Therefore we study these Efimov clusters for
 $n_0=0$ and $a_B=0$. In this case $\langle \hat{N} \rangle =\langle \sum_{\bm{k}} \hat{a}^{\dagger}_{\bm{k}} \hat{a}_{\bm{k}} \rangle=\langle \hat{N}_{ex} \rangle$. The polaron minimum disappears in this case, because the formation of the polaron is driven by the linear term in Eq.\ (\ref{eq:Hamiltonian}).

It has been shown for identical bosons that Efimov clusters with more and more particles get more and more strongly bound \cite{vonstecher:2011}. This can be explained using simple dimensional arguments \cite{naidon:2017}: the binding energy scales with $N^2$ and the kinetic energy with $N$. However, in the heteronuclear case, where the bosons interact with the impurity but not with each other,  this is not so obvious, because now also the binding energy scales with $N$. What happens in this case depends on the used models and parameters.

We show now that the same trend as in the homonuclear case also exists in our model for non-interacting bosons. Instead of exactly solving the $N$-body problem, for which the complexity grows exponentially with $N$, we take a variational Gaussian state (without coherent part), which consists of a superposition of states of even boson number. To obtain a variational bound on the energy for fixed $\langle \hat{N} \rangle$, we use imaginary time evolution \cite{shi:2017} to optimize the energy. To keep $\langle \hat{N} \rangle$ constant, we include a chemical potential which dynamically changes during the imaginary time evolution \cite{shi:2020}.

\begin{figure}[htp]
     \centering
     \includegraphics[width=0.48 \textwidth]{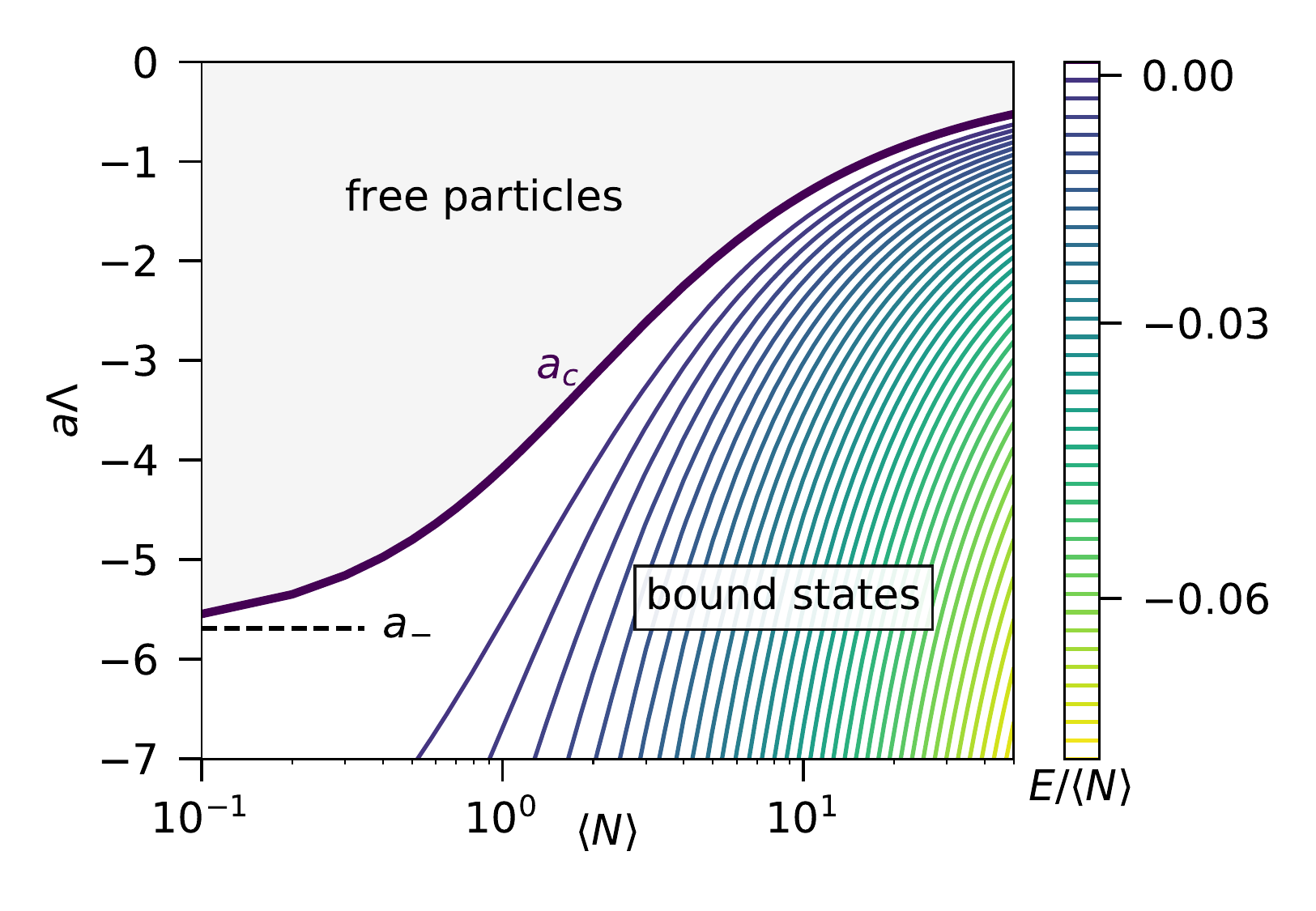}
     \caption{(Color online) Contour plot of the energy per particle $E/\langle N \rangle$ as a function of the expectation value of the number of particles $\langle N \rangle$ and the scattering length $a$ in terms of the three-body parameter $\Lambda$, for the mass ratio $M/m=6/133$. The energy is given in units of $\Lambda^2/M$. The bold line indicates the critical scattering length $a_c$ at which a bound state first appears.}
     \label{fig:coop_binding}
 \end{figure}

In Fig. \ref{fig:coop_binding} we plot the obtained energy per particle $E/\langle \hat{N} \rangle$ as a function of  $\langle \hat{N} \rangle$ and $a$ as a contour plot. We find that $|E|/\langle \hat{N} \rangle$ increases monotonically with $\langle \hat{N} \rangle$ and $|a|$. Furthermore, we consider the critical scattering length $a_c$ at which $E/\langle \hat{N} \rangle<0$, corresponding to the thick purple solid line in Fig.\ \ref{fig:coop_binding}. For $\langle \hat{N} \rangle \ll 1$, $a_c$  approaches $a_{-}\approx-5.7 \Lambda^{-1}$. In this regime, the Gaussian State can be approximated by:
\begin{equation}
      | \psi \rangle \approx (1+\frac{i}{2} \hat{\bm{\Psi}}^{\dagger} \Xi \hat{\bm{\Psi}}) |0 \rangle.
\end{equation}
In this limit, the energy of the state is entirely determined by the $N=2$ component and the Efimov effect is recovered. For increasing $\langle \hat{N} \rangle$, $|a_c|$ decreases, although it does not go all the way to zero \cite{ownpaper}. 

Altogether we see that the more particles the cluster contains, the more strongly it is bound and the smaller scattering length is needed for it to form. The impurity-mediated interaction between the bosons responsible for this ``cooperative binding" is described by the second term of the Hamiltonian in Eq.\ (\ref{eq:Hamiltonian}). This term originates from the Lee-Low-Pines transformation acting on the impurity momentum operator \cite{shchadilova:2016,ownpaper}. The driving force of the cluster formation can therefore be understood to be an ever more efficient usage of the kinetic energy of the impurity as the number of particles in the cluster grows.

Since the binding energy per particle increases monotonically with $N$, this means the total energy goes to $-\infty$ for an infinite number of particles. Therefore a many-body bound state of all particles in the system is the ground state of our Hamiltonian. Including interboson repulsion will push this many-body bound state up in energy and limit its number of particles. However, for weakly interacting bosons and a light impurity, such as considered here, this will only play an important role for large particle numbers. Moreover, Efimov clusters will be destroyed due to recombination effects before a large particle number can build up, so that their detailed microscopic structure is not relevant for our description of cold atomic gases.

\emph{Metastable polaron.-} We have now established that the ground state of the extended Fröhlich Hamiltonian is not a polaronic state, but a many-body bound state. However, to form such a many-body bound state dynamically, all these atoms would need to simultaneously come together into a small volume, which is extremely unlikely. It seems therefore likely that the polaron will survive as a metastable excited state of the Hamiltonian. This can be corroborated further by studying the mathematical structure of our Hamiltonian. In the limit of weak interactions,  $\langle \hat{N}_{ex} \rangle << 1$, and the Hamiltonian essentially reduces to the Fröhlich model, where the linear term in $\hat{b}^{\dagger}$ drives the polaron formation and the kinetic term (going with $\hat{b}^{\dagger} \hat{b}$)  counteracts this. This means that there will be a minimum of the energy as a function of particle number, such as displayed in Fig.\ \ref{fig:cartoon}. To study this picture quantitatively, we start by finding a local energy minimum on our variational manifold at small negative $a$ by imaginary time evolution. Indeed we find that this minimum corresponding to the polaron exists. We then update the position of the local minimum for incrementally increasing $|a|$. When the critical scattering length $a^{\ast}$ is reached, the local minimum turns into a saddle point, leading to a divergence of the particle number and the energy, indicating that the polaron has become unstable. This instability is caused by a coupling between the different particle number sectors. Once an Efimov cluster is formed, it will cascade down in energy into ever larger clusters.

\begin{figure}[tp]
    \centering
    \includegraphics[width=0.48 \textwidth]{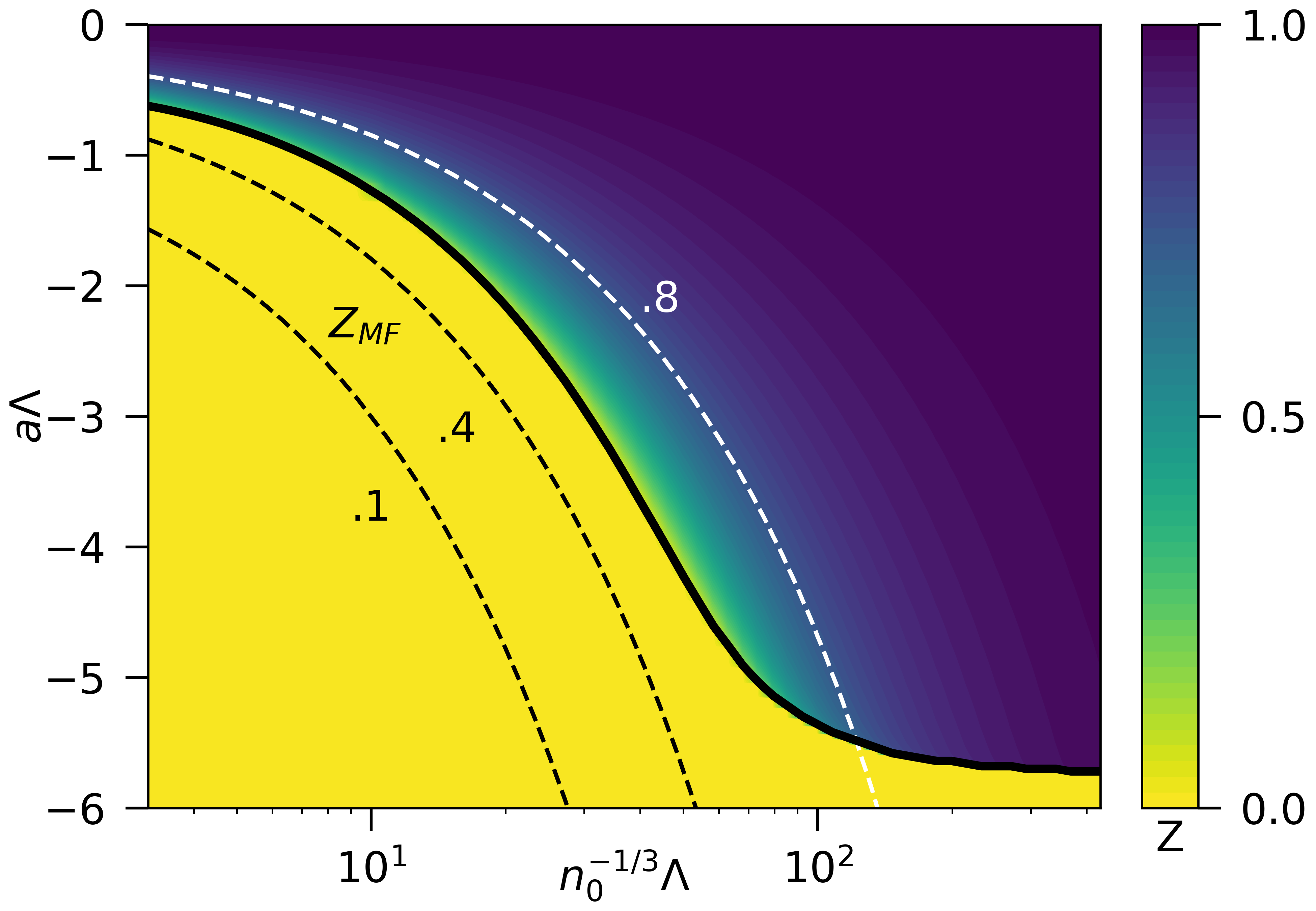}
    \caption{(Color online) The critical scattering length $a^{\ast}$ at which the polaron breaks down (black solid line) as a function of the mean interparticle distance $n_0^{-1/3} \Lambda$, for $M/m=6/133$ and $a_B\Lambda=0.1$, plotted together with a colormap of the quasiparticle weight $Z$. The dashed lines are contours of the mean field quasiparticle weight $Z_{MF}$. }
    \label{fig:ac_Z}
\end{figure}

In Fig.\ \ref{fig:ac_Z} we plot $a^{\ast}$ as a function of the average interbosonic distance $n_0^{-1/3}\Lambda$. We have used a small interboson scattering length $a_B=0.1 \Lambda^{-1}$, where we have verified that Bogoliubov approximation for the bosons is well-justified. This interboson repulsion serves to keep the size of the polaron cloud finite \cite{guenther:2021}. In the background of Fig.\ \ref{fig:ac_Z}, we plot the quasiparticle weight $Z=|\langle \psi | BEC \rangle|^2$  as a colormap.  The dashed lines are contours corresponding to the quasiparticle weight obtained from mean-field theory. We see that the quasiparticle weight is drastically affected by introducing correlations between the excitations, especially close to the instability, where it drops to zero. Surprisingly the polaronic energy only experiences a small shift ($<10\%$) compared to the mean field result \cite{ownpaper}.

\emph{Breakdown of polaron and shifted Efimov resonance.-} Let us now consider the black solid line in Fig.\ 3, corresponding to $a^{\ast}$. In the low density limit (\emph{i.e.}, for large $n_0^{-1/3}\Lambda$)  $a^{\ast}$ reduces to $a_{-}$. To understand this we need to consider the nature of our variational manifold. The breakdown of the polaron indicates that it is no longer stable against some perturbation of the wave function. For Gaussian states, the allowed perturbations are single and double excitations from the BEC. The instability of the polaron thus occurs when adding one or two excitations can trigger the onset of cluster formation. For an impurity in the vacuum, this instability corresponds exactly to the onset of Efimov state formation: the impurity can form a bound state when two particles are added.

In analogy to this scenario, the polaronic instability can be regarded as a many-body shifted Efimov resonance.
 When the background density is increased, $|a^{\ast}|$ decreases. Since the impurity already collected some particles in its polaron cloud, scattering of one or two additional particles on the polaron can now lead to formation of larger Efimov clusters. Due to the cooperative binding effect, this can already happen at scattering lengths smaller than $a_{-}$.
One fascinating aspect of this, is that the shift in the scattering length $a^{\ast}$ is continuous.
Where the average over particle numbers was purely classical in Fig.\ \ref{fig:coop_binding}, here the Hamiltonian coherently couples the different particle number sectors. This means that instead of getting a classical average over Efimov clusters of different particle numbers, we genuinely get a coherent superposition of those. This highlights an intriguing effect of coupling to a quantum mechanical medium such as a BEC.
 
 \begin{figure}[tp]
    \centering
    \includegraphics[width=0.48\textwidth]{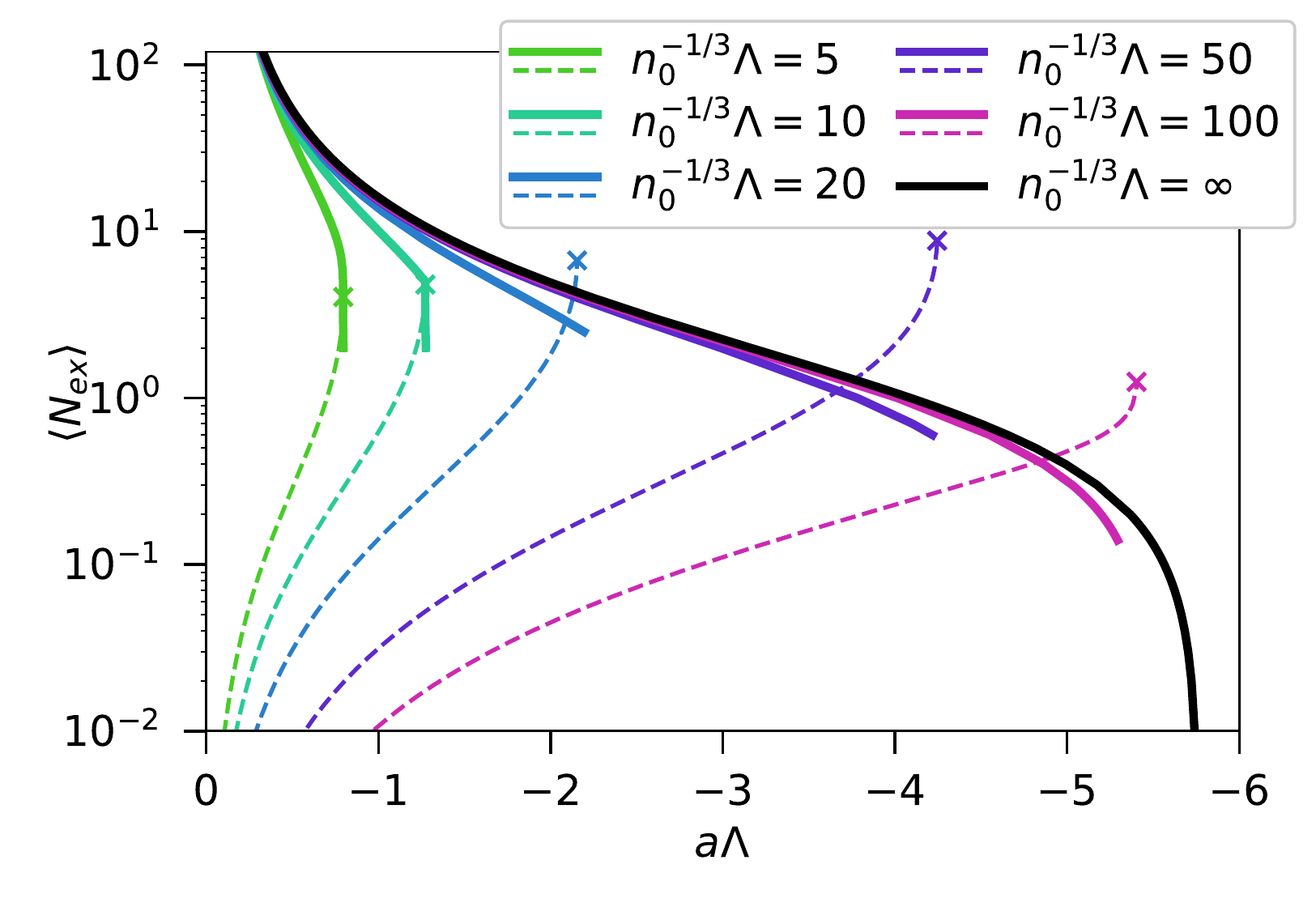}
    \caption{(Color online) The number of excitations in the polaron cloud $\langle \hat{N}_{ex} \rangle$ (dashed lines) as a function of scattering length $a\Lambda$ and density $n_0$ (indicated by color), and the number of excitations needed to form a bound state with lower energy than the polaron energy (solid lines).  The crosses indicate the points at which the polaron destabilizes. The interboson scattering length $a_B =0.1 \Lambda^{-1}$ and $M/m=6/133$. 
}
    \label{fig:polaron_partcnumb}
\end{figure}
 
 The mechanism of the polaronic instability is further elucidated in Fig.\ \ref{fig:polaron_partcnumb}. Here we plot the number of excitations $\langle \hat{N}_{ex} \rangle$ in the polaron cloud (dashed lines) and the number of excitations needed to form a many-body bound state lower in energy than the polaron (solid lines) as a function $a$ and $n_0$. The crosses at the end of the dashed lines indicate the points at which the polaron becomes unstable. The black solid line corresponds to the scattering length $a_c$ in Fig.\ \ref{fig:coop_binding} and the colored solid lines were calculated with the same method, but now including a finite background density $n_0$. Note that Fig.\ \ref{fig:polaron_partcnumb} is closely related to Fig.\ \ref{fig:cartoon}. The dashed lines in this figure correspond to the local minimum in Fig.\ \ref{fig:cartoon} (indicated by the black dot) and the solid lines correspond to the other side of the potential barrier at the same energy as the local minimum (indicated by the black square).
 
We find that the number of excitations in the polaron cloud increases as a function of $|a|$. For small $|a|$, the dashed lines lie below the solid lines, meaning that $\langle \hat{N}_{ex} \rangle$ is too small to facilitate many-body bound state formation and that the polaron is thus metastable. For small densities, the dashed lines cross the solid lines as $|a|$ increases. This is unexpected having Fig.\ \ref{fig:cartoon} in mind and this indicates that the 1D picture in Fig.\ \ref{fig:cartoon} is too simplistic. The explanation is that for small densities the extent of the polaron cloud, determined by the healing length of the BEC, is much larger than the size of the Efimov states. So even though the bound state formation would be possible with the given number of particles, most of the particles are too distant to contribute to the bound state formation. For larger densities and thus smaller healing lengths, the polaron is comparable in size or smaller than the Efimov state. In this case the bound state is formed as soon as the polaron contains enough particles, which means that the instability occurs exactly when the dashed line hits the solid line. As discussed before, we find that $|a^{\ast}|$ becomes smaller for larger densities. From Fig.\ 4 we see that this is mainly due to an increase in $\langle \hat{N}_{ex} \rangle$ (upward shift of the dashed lines), but also partially by the stabilization of the bound states due to the linear Fröhlich term (small downward shift of solid lines).

\emph{Experimental observation.-} To observe our finding experimentally, we propose to use light impurities such as Li in a BEC of K \cite{huang:2019}, Rb \cite{maier:2015} or Cs \cite{tung:2014,pires:2014,desalvo:2019}, where the Efimov effect is most prominent.
Our results could be observed in three-body loss measurements, similar to how the Efimov effect \cite{tung:2014,pires:2014} was observed, but performed in a BEC instead of a dilute thermal cloud. After adiabatic preparation of the polaron state by a ramp of the magnetic field, crossing the instability of the polaron should lead to a drastic enhancement of the loss through recombination. This loss enhancement feature could then be measured as a function of the density. Injection spectroscopy  \cite{jorgensen:2016,hu:2016,yan:2019} could complement loss measurements, since the strong decrease in quasiparticle weight close to the breakdown of the polaron should lead to a strong broadening of the polaronic spectral line. In injection spectroscopy also excited Efimov states could be observed  when favorable mass ratios are used \cite{sun:2017}. 

\emph{Conclusion.-} We have explored how immersion in a quantum medium can modify an impurity's properties and chemical reactivity. In particular, we have studied the connection between Efimov physics and polaron formation in case of light impurities in a BEC. Using a variational method based on Gaussian States in the reference frame of the impurity, we have shown that the interplay of these two effects leads to an instability of the polaron towards Efimov cluster formation. This instability can be interpreted as a many-body shifted Efimov resonance, which can be probed experimentally through radiofrequency spectrocopy and the chemical reactions of three-body recombination.

Future directions include the real-time dynamics of the polaron \cite{drescher:2018} with Gaussian States to answer key open questions about the competition between few-body recombination and formation of many-body correlated states. Related venues include competition between Auger recombination of excitons in semiconductors and formation of strongly correlated exciton liquids. Furthermore, also the bipolaron problem \cite{camacho:2018,naidon:2018,panochko:2021}, involving medium-induced polaron-polaron interactions, is of interest and our results may provide new insight into the role three-body correlations play in many-body systems such as quenched BECs \cite{makotyn:2014,piatecki:2014,eismann:2016,klauss:2017,colussi:2018,eigen:2018,incao:2018,colussi:2020,musolino:2021,ownpaper}.

\emph{Acknowledgements.-} The authors are thankful to Tommaso Guaita and Tao Shi for useful discussions. J.I.C. acknowledges funding from ERC Advanced Grant QENOCOBA under the EU Horizon
2020 program (Grant Agreement No. 742102). R.S. acknowledges support from the Deutsche Forschungsgemeinschaft (DFG,
German Research Foundation) under Germany’s Excellence Strategy–EXC–2111–390814868.

\bibliography{Paper.bib}

\end{document}